\documentclass[aps,pra,twocolumn,showpacs,groupedaddress,letterpaper,nofootinbib]{revtex4-1}

\usepackage{graphicx}
\usepackage{subfigure}
\usepackage{color}
\usepackage{physymb}
\usepackage{enumitem}
\usepackage{mdframed}
\usepackage{framed}
\usepackage{epstopdf}
\usepackage{natbib}
\usepackage{hyperref}
\usepackage{url}
\usepackage{multirow}

\begin{document}

\title{Ontological Flexibility and the Learning of Quantum Mechanics}


\author{Charles Baily}
\affiliation{School of Physics and Astronomy, University of St Andrews, St Andrews, Fife KY16 9SS Scotland, UK}

\author{Noah D. Finkelstein}
\affiliation{Department of Physics, University of Colorado, 390 UCB, Boulder, CO 80309 USA}

\begin{abstract}
One area of physics education research has focused on the nature of ontologies (mental categorizations of concepts, substances and processes), and how they might be used to gain insight into student thinking when learning classical physics.  There has been some debate about whether student and expert ontologies in classical contexts should be thought of as stable cognitive structures or dynamic cognitive processes, and what implications these different perspectives have for instruction.  We extend this discussion of ontologies and their role in learning and cognition to the context of quantum physics, by first considering various types of epistemological and ontological commitments exhibited by experts in their descriptions of quantum phenomena.  Excerpts from student interviews and responses to survey questions are used to demonstrate the contextual nature of students' quantum ontologies, which can be best understood within a \textit{resources} framework.  Instructional implications are discussed, including a brief description of a modern physics curriculum designed to explicitly attend to student ontologies.
\end{abstract}

\maketitle

\section{Introduction}

Physics education research has shown that, when it comes to certain topics, student thinking can be notoriously resistant to change via direct instruction.  In many cases, such learning difficulties can be understood as stemming from the \textit{prior knowledge} of students, which somehow interferes with the learning process, so that something must change before a proper (scientifically normative) understanding can be achieved.  Precisely what it is that is supposed to change in the minds of learners, whether it be concepts, beliefs, epistemological framings or ontologies, is where education researchers primarily diverge \cite{disessa2006concept}.

An \textit{ontology} is normally understood within the learning sciences as a mental categorization of concepts, substances and processes according to their fundamental properties, where entities with similar characteristics belong to the same or similar categories.  One line of research posits that some of the conceptual barriers faced by students can be traced to inappropriate degrees of commitment to ontological category assignments that are unproductive.  A generally accepted hypothesis is that when learners encounter an unfamiliar concept, they engage in a (conscious or unconscious) act of ontological categorization, whereby the new concept is sorted according to whatever information is available at the time.  This may include: the context in which the concept is introduced, its co-occurence with other concepts, or language patterns that are indicative of its ontological nature.  Once a category has been settled upon, it is thought that learners then automatically associate with that concept the attributes of other, more familiar concepts that fall within the same category.  In other words, the new concept \textit{inherits} the characteristics of other concepts that are, in the mind of the learner, ontologically similar \cite{slotta2011defense}.

There are opposing views with respect to ontologies about the nature of novice and expert reasoning, each with different implications for instruction.  According to some, when a learner's category assignment for a given concept is sufficiently distinct from the targeted (scientifically accepted) category, the process of reassignment cannot come about in gradual steps, and the initial conceptualization must be suppressed in favor of one with other attributes \cite{chi2005common,slotta2011defense}.  This \textit{incompatibility hypothesis} is a key aspect of Chi's description of radical conceptual change, and motivates an instructional approach that adheres to `ontological correctness'.  For example, \textit{emergent processes} (such as electric current, resulting from the net motion of individual charged particles) are often alternatively conceptualized by students as \textit{material substances} (e.g., electric current as a fluid that can be stored and consumed) \cite{reiner2000reason}. Slotta and Chi suggest that such misconceptions might be avoided if instructors were to eschew materialistic analogies for electricity (such as water flowing through a pipe) \cite{slotta2006ontology}.

Others object to this delineation of ontologies into distinct, normative categories, and instead argue that both students and experts often bridge between (and sometimes blend) different ontologies in ways that are productive \cite{gupta2010dynamic,hammer2011intuitive}.  Examples have been given of students productively using substance-based ontologies for energy \cite{scherr2012energy} and gravity \cite{gupta2014gravity} as motivation for an instructional approach that leverages exactly the type of novice reasoning that Slotta and Chi suggest ought to be suppressed.  Gupta, \textit{et al.} argue that, rather than there being just one ``correct'' ontology for a given concept that should be promoted in the classroom, different ontologies can be sometimes more productive, and sometimes less, depending on the specific context. \cite{gupta2010dynamic}.

The differences between these two models of cognition are analogous to those between material substances and emergent processes: ontologies as stable cognitive structures (which is one way of accounting for the robustness of student misconceptions) versus ontologies as dynamic cognitive processes, emerging in real time through the coordinated activation of conceptual resources (that are in themselves neither right nor wrong) \cite{hammer2005resources}.  Although they differ in their implications for instruction, these two perspectives are not entirely incompatible.  Both agree that the learning of new concepts is mediated (and sometimes hindered) by prior knowledge, and that difficulties can arise from the misattribution of ontological characteristics to unfamiliar concepts.  Most importantly, both allow for the possibility of productively using multiple \textit{parallel} ontologies for a single concept, \textit{if} accompanied by a sufficient understanding of the inherent bounds and limitations of each.

We wish to extend this discussion on learning and cognition to the context of quantum physics, where students often have difficulty reconciling their classical intuitions with the behavior of quantum entities.  We argue that some of these difficulties can be understood in terms of what we'll call \textit{classical attribute inheritance}.  For example, many students persist in always ascribing a localized position to electrons because they also possess the attributes of mass and charge, which are all normally associated with classical particles.  This may then act as a barrier to developing a full understanding of phenomena that entail both particle and wave descriptions.

Our prior research has demonstrated that student thinking can be differentially influenced by the myriad ways in which instructors choose (or choose not) to address interpretive themes in quantum mechanics, and that these instructional choices manifest themselves both explicitly and implicitly in the classroom \cite{baily2009development,baily2010teaching,baily2010refined,baily2010hidden,baily2011thesis,baily2015modern}.  In this paper, we show that students exhibit varying degrees of \textit{flexibility} in their ontological categorizations of electrons, and present evidence of them not only switching between categories (both within and across contexts), but also creating a blended category by classifying electrons as simultaneously both particle and wave.  We find that students frequently modify their conceptions of quanta in a piecewise manner, often without looking for or requiring internal consistency.  Even when their instructors de-emphasize interpretation (explicitly or otherwise), students still develop their own ideas about quantum phenomena, some of which emerge spontaneously as a form of sense making.  Such insights have motivated our development of a curriculum designed to strengthen students' abilities to physically interpret quantum theory, and to understand the domains of applicability of those interpretations.

\section{\label{sec:QMOntology}Ontologies in Quantum Physics}

\begin{figure}
    \includegraphics[height=0.8in, width=3.5in]{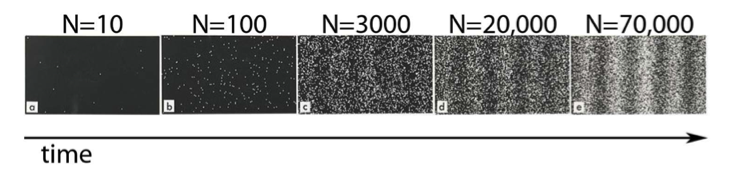}
\caption{Buildup of an electron interference pattern.  Single electrons are initially detected at seemingly random places, yet an interference pattern is still observed after detecting many electrons. \cite{tonomura1989electron}}\label{fig:electronbuildup}
\end{figure}

Wave-particle duality makes ontological flexibility \textit{necessary} for understanding the various ways that physicists describe quantum phenomena.  For example, when a double-slit experiment is performed with a low-intensity beam, each electron will register individually at the detector, yet an interference pattern will still be seen to develop over time \cite{tonomura1989electron,frabboni2012buildup}. [See Fig.\ \ref{fig:electronbuildup}.]  Interference is a property associated with waves, whereas localized detections are indicative of a particle-like nature.  Expert physicists will interpret this experimental result differently, depending on their epistemological and ontological commitments.

The standard \textit{Copenhagen} interpretation of quantum mechanics \cite{faye2008copenhagen} would say this experiment reveals two sides of a more abstract whole; an electron is neither particle nor wave.  The dual use of (classically) distinct ontological categories is just a way of understanding the behavior of electrons in terms of more familiar macroscopic concepts.  A wave function is used to describe electrons as they propagate through space, and the \textit{collapse postulate} is invoked to explain localized detections, but any switch between `particle' and `wave' occurs only in the sense of how the electron is being represented.  The wave function is nothing more than a mathematical construct used to make predictions about measurement outcomes, without reference to any underlying reality.

From a \textit{Matter-Wave} perspective, the wave function is (for all intents and purposes) physically real: each electron \textit{is} a delocalized wave as it propagates through both slits and interferes with itself; it then randomly deposits its energy at a single point in space when it interacts with the detector.  The \textit{collapse of the wave function} is viewed as a process not described by the Schr\"{o}dinger equation, in which the electron physically transitions from a delocalized state (wave) to one that is localized in space (particle) \cite{bassi2013collapse}.  This way of thinking about electrons \textit{might} be at odds with relativity theory, though this is generally only problematic within the context of distant, correlated measurements performed on entangled systems \cite{maudlin2011relative}.  Many physicists are comfortable with employing this model in situations where relativity does not come into play, which underscores the fact that not all experts feel the need to subscribe to a single ontology that has universal applicability.

This experiment might also be used to motivate a \textit{Statistical} (or, ensemble) interpretation of quantum mechanics \cite{ballentine1970statistical}, as Ballentine does in the introductory chapter of his graduate-level textbook: \begin{quote}``When first discovered, particle diffraction was a source of great puzzlement.  Are `particles' really `waves'?  In the early experiments, the diffraction patterns were detected holistically by means of a photographic plate, which could not detect individual particles.  As a result, the notion grew that particle and wave properties were mutually incompatible, or \textit{complementary}, in the sense that different measurement apparatuses would be required to observe them.  That idea, however, was only an unfortunate generalization from a technological limitation.  Today it is possible to detect the arrival of individual electrons, and to see the diffraction pattern emerge as a statistical pattern made up of many small spots.  Evidently, quantum particles are indeed particles, but particles whose behavior is very different from what classical physics would have led us to expect.'' \cite{ballentine1998modern}\end{quote} Ballentine \textit{assumes} that localized detections imply that electrons are localized throughout the experiment, always passing through one slit or the other (but not both).  He explains diffraction patterns in terms of a quantized transfer of momentum between a localized particle and a periodic object.  An electron may be different from a classical particle, but it is a particle nonetheless, and departures from classical physics can mostly be ascribed to the existence of a quantum of action.  The wave function only encodes probabilities for the outcomes of measurements performed on an ensemble of identically prepared systems.

\begin{figure}
    \includegraphics[height=2.1in, width=3.2in]{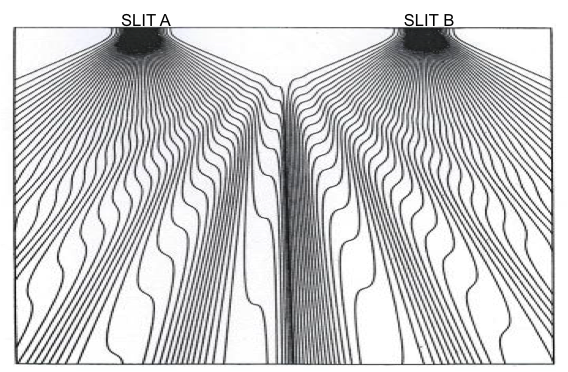}
\caption{Time-averaged trajectories of localized particles in a double-slit experiment, according to Bohm's pilot-wave interpretation of quantum mechanics. \cite{bohm1995undivided}}\label{fig:bohm-doubleslit}
\end{figure}

In a \textit{Pilot-Wave} theory \cite{bohm1995undivided,norsen2013pilotwave}, localized particles follow trajectories determined by a quantum potential that interacts non-locally with the apparatus; an electron is simultaneously both a particle that goes through only one slit, and a wave that passes through both.  The initial conditions for each electron (which are unknowable to the experimenter) predetermine where its particle aspect will land on the detecting screen. [See Fig.\ \ref{fig:bohm-doubleslit}.]  This hidden-variable theory is not incompatible with the results of Bell's theorem, which only requires such theories to be non-local in order to reproduce the predictions of quantum mechanics \cite{bell1964epr}.

There are still more ways to interpret these experimental results \cite{wallace2012everett,fuchs2014qb}, but they are typically not discussed in any detail in undergraduate quantum mechanics instruction.  The term \textit{Agnostic} can be used to describe a perspective that takes no firm stance on the correctness of any one of the above interpretations.  The Copenhagen interpretation seems to be favored by most physicists, not only for historical reasons; by making no claims about the true nature of reality (other than that it is inaccessible), it allows physicists to focus more on the predictive power of quantum mechanics (``Shut up and calculate!'' \cite{mermin1989pillow}).  This pragmatic attitude is often reflected in the classroom, and so we also use the terms Copenhagen/Agnostic jointly below to denote a common instructional approach that de-emphasizes interpretive discussions in favor of developing mathematical tools.

Regardless of one's personal preferences, it should be clear that a thorough understanding of quantum theory requires some amount of ontological flexibility.  Even though Ballentine may appear to be inflexible in his ontological views on quantum particles, he must still have facility with other ontologies in order to properly understand the views of other physicists.  Experts have developed the sophistication to know when it is (and is not) appropriate to employ each type of description.  When first learning about quantum mechanics, students are in the process of developing this sophistication, and they can be influenced by the pedagogical choices of their instructors.

\section{\label{sec:StudentOntologies}Students' Quantum Ontologies}

\subsection{\label{survey}Online Surveys and Aggregate Responses}

We first observed the contextual nature of student ontologies in the responses of modern physics students at the University of Colorado to a series of survey questions designed to probe their epistemological and ontological commitments in the context of quantum mechanics.  It was emphasized at the beginning of this online survey that we were asking students to express their own beliefs, and that their specific answers would not affect any evaluation of them as students.  One of the survey questions asked them to respond using a 5-point Likert scale (from \textit{strong agreement} to \textit{strong disagreement}) to the statement, \textit{When not being observed, an electron in an atom still exists at a definite (but unknown) position at each moment in time}, and to explain their reasoning.

Some of the survey statements have evolved over time, primarily in the early stages of our research; modifications were usually motivated by a fair number of students providing reasoning that indicated they were not interpreting the statements as intended.  We conducted validation interviews with 19 students in 2009 (see below), after which the phrasing has remained essentially unchanged.  The survey responses presented in this paper were all collected after the validation interviews had taken place.  The \textit{agree} and \textit{strongly agree} responses have been collapsed into a single category (agreement), and similarly for \textit{disagree} and \textit{strongly disagree}.

\begin{figure}
    \includegraphics[height=1.5in, width=3.5in]{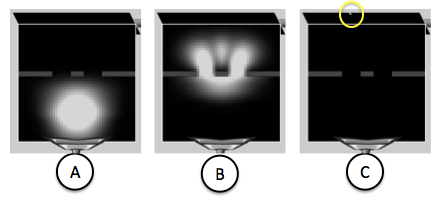}
\caption{A sequence of screen shots from the Quantum Wave Interference PhET simulation \cite{phetqwi}: (A) a bright spot emerges from an electron gun; (B) passes through both slits; and (C) a single electron is detected on the far screen (highlighted in this figure by the circle). After many electrons, a fringe pattern develops (not shown).}\label{fig:phetqwi}
\end{figure}

An additional essay question at the end of the survey presented statements made by three \textit{fictional} students regarding their interpretation of how the double-slit experiment with single electrons is depicted in the PhET Quantum Wave Interference simulation \cite{phetqwi} (as shown in Fig. \ref{fig:phetqwi}):

\begin{quote}
\textbf{Student 1:} The probability density is so large because we don't know the true position of the electron. Since only a single dot at a time appears on the detecting screen, the electron must have been a tiny particle, traveling somewhere inside that blob, so that the electron went through one slit or the other on its way to the point where it was detected.

\textbf{Student 2:} The blob represents the electron itself, since an electron is described by a wave packet that will spread out over time. The electron acts as a wave and will go
through both slits and interfere with itself. That's why a distinct interference pattern will show up on the screen after shooting many electrons.

\textbf{Student 3:} Quantum mechanics is only about predicting the outcomes of measurements, so we really can't know anything about what the electron is doing between being emitted from the gun and being detected on the screen.
\end{quote}  Respondents were asked to state which students (if any) they agreed with, and to explain their reasoning.

Student responses to the double-slit essay question were generally aligned with the type of instruction they received in the classroom \cite{baily2015modern}.  One instructor taught that each electron goes through one slit or the other, but any attempt to determine which one will disrupt the interference pattern; these students were the most likely to agree with Student 1.  Student 2's statement was overwhelmingly preferred by students from a class where the instructor described the electron as a wave that passes through both slits.  Responses were more varied from a course in which the instructor explained that a ``quantum mechanical wave of probability'' passes through both slits, but ultimately emphasized calculating features of the interference pattern over physically interpreting the result; these students were equally likely (within statistical error) to agree with any one of the three statements.

We have observed a common tendency for instructors to spend more time discussing the physical interpretation of the wave function at the outset of the course (e.g. when introducing the double-slit experiment), and devote much less time to such explicit discussion in later topics (e.g., the Schr\"{o}dinger model of hydrogen).  In contrast to the results for the double-slit essay question, most students from all three types of courses agreed with the statement that atomic electrons exist as localized particles when not being observed. [See Ref. \cite{baily2015modern}, where the influence of instruction on student thinking is explored in greater detail.]

\begin{figure}[b]
    \includegraphics[height=2.1in, width=3.2in]{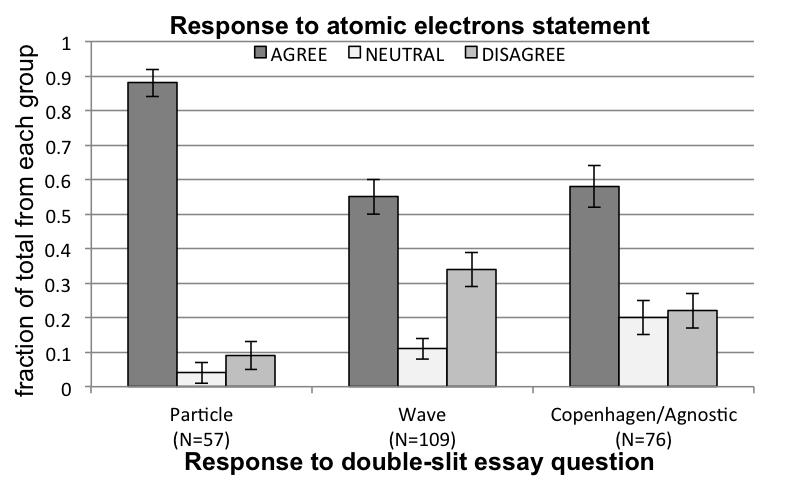}
\caption{Combined student responses to the statement: \textit{When not being observed, an electron in an atom still exists at a definite (but unknown) position at each moment in time}, grouped by how those same students responded to the double-slit essay question.  Error bars represent the standard error on the proportion.}\label{fig:doubleslit-atom}
\end{figure}

Combining student data from all three of these courses, so that responses to the statement about atomic electrons are grouped by how those same students responded to the double-slit essay question [Fig.\ \ref{fig:doubleslit-atom}], we see that students who preferred the particle description of electrons in the double-slit experiment were the most consistent, in that almost all of them also agreed with the statement that atomic electrons are localized in space.  Over half of those who preferred a wave description of electrons in the double-slit experiment agreed that atomic electrons exist as localized particles, and a third of them disagreed.  Students who preferred the Copenhagen/Agnostic statement of fictional Student 3 also tended to agree with this statement; those who did not agree were evenly split between disagreement and a neutral stance.

Such findings indicate that students can be influenced by explicit instruction in a measurable way, but that their ontological perspectives on electrons are not necessarily stable across contexts. Notice also that most students defaulted to an intuitive, classical view of electrons as localized particles in a context where instructors did not adequately attend to student ontologies.  These results motivated a more detailed exploration of student conceptions of quantum phenomena.

\subsection{\label{method}Student Interviews: Participants and Methods}

Students were recruited for 1-hour interviews from four modern physics offerings at the University of Colorado in a single academic year (two offerings per semester, one for engineering students and one for physics majors).  A total of 19 students were interviewed, either in the last week of the semester or after the course had ended.  Participants from the physics courses were all physics or engineering physics majors, plus one astronomy major; those from the engineering courses were all mechanical or electrical engineering majors, plus one mathematics major.  The average final course grade for all 19 students was 3.4 out of 4.0, while overall averages for each course fell in the 2.0 {-} 3.0 range, meaning the participants were generally better than average students (as might be expected for a group of volunteers).

12 of the 19 students had already responded to the post-instruction online survey before the interviews took place; their survey responses were consistent with their interview responses in almost every case (two of them switched from a neutral stance in the survey to agreement during the interview with the atomic electrons statement).  We found no correlations between how students responded during the interviews and how their particular instructor discussed interpretation during class, likely due to the small number of participants from each course.

Each interview began by asking the student to simply describe an electron, in words or pictures.  15 of the 19 students talked about them first and foremost as constituents of atoms, and all but one of the remaining four eventually mentioned atomic electrons without any prompting.  Most of them used a planetary model as a first-pass description, though every student claimed (when asked) that they were aware the Bohr model is not an accurate description.  When asked to elaborate, the majority of them eventually stated that the electron is more properly described by an electron cloud, or a cloud of probability.  Once a clear picture had been established of how each student thought of electrons in the context of atoms, they were then asked to respond to the atomic electrons statement and explain their reasoning.  In every case, their responses were consistent with the descriptions they had given immediately prior.

We then asked students to describe the setup of the double-slit experiment and what is observed, so as to first establish that each of them had sufficient content knowledge before going on to discuss the implications of the results.  Every student talked about the fringe pattern in terms of wave interference, and all but one knew that attempts to determine which-path information would disrupt the pattern.  They were also all aware that the experiment could be run with single electrons, and that an interference pattern would still develop over time.

Afterwards, students responded to the double-slit essay question by reading the three statements one at a time, discussing their response to each before moving on to the next statement.  The interviewees did not necessarily respond to these statements in a way that was consistent with their earlier view of atomic electrons.  Some switched from particle to wave descriptions, though no student applied first a wave description of atomic electrons and then a particle description in the double-slit experiment.  In some cases, their responses to one of the three statements was inconsistent with their responses to the other two.

\subsection{\label{sec:interview}Individual Survey and Interview Responses}

\subsubsection{\label{sec:inherit}Particle ontologies and classical attribute inheritance}

In classical physics, as in colloquial usage, the word \textit{particle} connotes some small object with negligible spatial extent.  It should therefore not be too surprising that many introductory quantum physics students persist in thinking of electrons in this way, particularly when instructors continue to refer to them as particles, even after they have stressed that they also possess distinctly different attributes.

During the interviews, three students remained committed throughout to a strictly particle description of electrons.  They all claimed that wave functions are just mathematical tools, used only for describing where an electron is likely to be found when a position measurement is made.  Students A \& B (among others; see below) also made statements that suggest their ontological views had been influenced by classical attribute inheritance: spatial localization goes hand-in-hand with other attributes like mass and charge:

\begin{quote}\textbf{STUDENT A:} I guess an electron has to [exist at] a definite point. It is a particle, we've found it has mass and it has these intrinsic qualities, like the charge it has, so it will have a definite position, but due to uncertainty it will be a position that is unknown.\end{quote}

\begin{quote}\textbf{STUDENT B:} I can't picture it any other way.  I can't picture it not having a definite position, because it is a piece of mass, too.\end{quote}

Similar statements have been made by other students who were not interviewed, in their online responses to the atomic electrons statement; for example:

\begin{quote}\textbf{STUDENT C:} I feel that electrons are particles with mass so they will have a definite position.\end{quote}

\begin{quote}\textbf{STUDENT D:} I believe even if it's not being observed, it still produces the negative charge.  Since charge doesn't just go away, the electron MUST exist there.\end{quote}

What is important to notice about these statements is the difficulty students have with attributing mass or charge to an electron without also ascribing to it a localized position.  Introductory quantum physics students know that classical electrons have all three attributes, but they are less comfortable with the idea of a quantum electron that sometimes possesses only two of the three.  In a similar vein, others will argue that charge (or mass) could not be conserved if the electron were not located at a single position.  These student quotes serve as evidence for how the ontology being employed by students may influence their reasoning (or vice-versa), and how prior knowledge can interfere with developing new conceptions of light and matter.

\subsubsection{\label{sec:intuition}Conflicts with intuition}

Four other students initially described electrons as localized particles during the interviews, but eventually wavered in their commitment, ultimately distinguishing between their intuitive conceptions (what made personal sense to them) and what they had learned from instruction.  For example, Student E agreed with the notion of localized atomic electrons, and continued with this line of thinking as he first described the double-slit experiment.  The following excerpt begins as he was reflecting on the statement made by fictional Student 1:

\begin{quote}\textbf{STUDENT E:} I would agree with what [the first fictional student] is saying, that the electron is traveling somewhere inside that probability density blob, and it is a tiny particle.  The problem here that I see is that [the fictional student is saying] the electron went through one slit or the other. [PAUSE] So, now I'm disagreeing with myself.  OK, my intuition is fighting me right now.  I said earlier that there should be one point in here that is the electron, and it goes through here and hits the screen, but I also know that I've been told that the electron goes through both slits and that's what gives you the interference pattern.  Interesting. [LONG PAUSE] OK, somehow I feel like the answer is going to be that this probability density, it is the electron, and that can go through both slits, and then when it's observed with this screen, the probability density wave collapses, and then only exists at one point.  But at the same time I feel that there should be a single particle, and that somehow a single, finite particle exists in this wave, and will either travel through one slit or the other.  Why that particle would be affected by this slit, like its direction, like why would it go here or here, why would a single particle be affected by a slit?  That I don't have an answer to, other than that it's the wave that's actually being propagated, the wave is the electron.\end{quote}

Regardless of what Student E may have learned during the semester, his post-instruction thinking was initially dominated by a particle ontology for electrons, regardless of the context.  It is not that he hadn't learned to think about the double-slit experiment in terms of waves; in fact, he had originally explained the fringe pattern as resulting from wave interference.  The association of the wave with the electron itself was not cued until he was confronted with the notion that each electron had passed through a single slit.  If the electron were always localized, then this would be the only logical conclusion, but he could not reconcile this with the experimental results.  Perhaps this student would have appealed to Ballentine's explanation for electron diffraction had he had known of it, but he was still capable of switching to a more productive ontology when the need arose, even though it was at odds with his intuition.  We would argue, however, that it would be preferable for students to avoid cognitive dissonance when deciding what type of ontology to employ.

\subsubsection{\label{sec:copenhagen}Copenhagen-like reasoning}

Student F was one of three students who were explicit about not thinking of an electron as either particle or wave, but recognized that one description may be more productive than another in particular situations.  Just before being asked to respond to the double-slit essay question, he stated:

\begin{quote}\textbf{STUDENT F:} I don't really have much of a conception of how it exists from here where we emit it to the wall where it impacts.  You can't think of it as a particle, cause that doesn't make any sense, and thinking of something that you know has a mass as a wave doesn't make much sense either.  I don't really have much of a feeling of what it is between [emission and detection], I just know that it's an electron, because I shot an electron out and it hit the wall, so I have to assume there was one in the middle.  We can conceive of it as this wave function, as this sort of smeared out thing of probability, but since we can't observe that, then I don't really see the point.\end{quote}

\noindent Notice that, similar to Students A$-$D, Student F says that it ``doesn't make much sense'' to associate wave characteristics with a entity that possesses mass.  Later, after reading the statement from fictional Student 1, he continued:

\begin{quote}\textbf{F:} Like I said before, I don't really see the need for a conception of the electron between [emission and detection], because we're not going to be able to observe it without changing it, so I disagree that it's a particle. Well, it is something that acts like a particle somewhere within that blob and it would technically have to go through one slit or the other if we think of it as a particle, but every time we try to observe that or verify it, it ruins the experiment and produces another kind of interference pattern, so you can't know that it's a particle going through one slit or the other, so I don't know why it would be helpful to think of it like that.

\textbf{INTERVIEWER:} So, if you were in class and a student were to make this statement, would you try to argue against it, or would you just say that it's not scientific to speculate if you can't observe it?

\textbf{F:} Well, I would sort of argue against it, but I don't know how you can [argue that] it's a bunch of tiny particles somewhere in here, and that's going to come up with an interference pattern.  I don't know how you would make sense of the interference if you thought of it as particles.  If you thought of it as particles that continued to obey this strange probability cloud, then I guess you could go with it.  But I think if somebody next to me said, yeah an electron is a particle so it goes through one slit or the other, but we don't know which slit it's going through and we won't...  Again, we're not going to see it, so it's not that it's actually wrong, but I don't know that it would lead constructively to understanding why interference happens.\end{quote}

Student F disagrees with thinking of the electron as a particle in the double-slit experiment, though he isn't willing to go so far as to say that it is ``wrong'' to do so.  Nor is he comfortable with describing it as a wave, if that can't be directly observed, but he concedes that the wave description is more \textit{productive} for explaining the fringe pattern.

\subsubsection{\label{sec:pilot}Pilot-Wave ontologies}

All of the instructors considered in our studies were explicit about particle and wave descriptions of electrons being mutually exclusive; they may exhibit characteristics of one or the other, but not both at the same moment in time.  Nor did they mention any type of Pilot-Wave interpretation of the double-slit experiment, though we are reminded that in-class instruction is not the only source of information about quantum physics, and that students are also capable of constructing their own ideas.  Three of the interviewees were fairly consistent in talking about electrons as being simultaneously a particle and a wave, such as Student G, whose description of the double-slit experiment is remarkably reminiscent of Fig. \ref{fig:bohm-doubleslit}.  Here, he is responding to the statement made by fictional Student 2:

\begin{quote}\textbf{STUDENT G:} `The electron acts as a wave and goes through both slits...'  Yeah and no, I think it's the probability, the possible paths, it seems like the probable paths for the electron to follow interact with themselves, but the electron itself follows just one of those paths.  So I don't think the electron goes through both slits.  I think that the possible...  It's like the electron rides on a track, like a rail, like a train rides on a rail, but those rails or tracks go through both slits, and the possible paths for the electrons to follow interfere with themselves, create the interference pattern, but the physical electron just rides on the tracks, it picks one.  Or maybe switches paths, if two of them cross.  I don't know, it seems that the electron has to be on one of those tracks, but the tracks themselves cause the interference pattern.\end{quote}

\noindent Not all students who viewed electrons as both particle and wave were as imaginative in their thinking, but this type of explanation have also been volunteered by students in their written responses to the double-slit essay question:

\begin{quote}\textbf{STUDENT H:} I personally visualize the situation as a flow of some fluid that travels through the two slits in waves.  It appears through all space as soon as the electron is fired.  The electron then rides this chaotic fluid toward the screen and strikes in a location that is somewhat determined by the interference patterns of the fluid.  Trying to measure this fluid flow collapses the waves created.\end{quote}

\subsubsection{\label{sec:matter}Matter-Wave ontologies}

Six of the interviewed students bridged between particle and wave descriptions of electrons in the double-slit experiment, such as Student I, who had just finished reading the statement by fictional Student 1:

\begin{quote}\textbf{STUDENT I:} No, I don't agree with that.  The electron went through both slits, it stopped being this blob the moment it hit the screen, but before we detected it, it was that blob.  That was its probability distribution function.  That electron went through both those slits, and it only stopped having that large probability function when we detected it.  So yeah, I would say that's false.

\textbf{INTERVIEWER:}  What would you say if you were sitting in class, and a student had this argument that we just detect a single dot in a specific case, and that makes me think that an electron is a point particle?

\textbf{I:}  Well, that's the whole wave-particle duality.  Like I said, it is one particle when we measure it, and we determine it is one particle, but until we measure it, it is this probability density function, and that electron did go through both the slits.  The probability was split, and that's why we pick up this interference pattern, right?  Yeah, so I pick up the interference pattern because the electron, we wouldn't have picked up the interference pattern if there was one slit, the electron would have to go through both for there to be an interference pattern, or else it would just be the same as if going through one slit, or a 50\% chance of it going through one slit and 50\% chance...  We wouldn't observe this interference pattern if it didn't go through both, it would have to go through both.

\textbf{INT:}  You had said that this [blob] is the electron.  What do you mean by that?

\textbf{I:} Well, it's hard to conceptualize.  For the probability density, in a way it's the electron, I mean it's the electron's wave function, which I would say is just as valid a part of the electron, you know, as anything else we know about the electron.  When that probability function splits, the electron, and goes through both these slits, so does the electron.\end{quote}

Student I initially alternated between talking about the electron going through both slits, and saying it was the probability density that went through both.  Only later did he clarify that, for him, they are essentially one and the same.

Finally, although Student J claimed to have accepted the Matter-Wave perspective as a legitimate way of describing an electron, he still maintained a sophisticated Agnostic stance:

\begin{quote}\textbf{STUDENT J:} The way I think of an electron, I cannot ascribe to it any definite position, definite but unknown position.  I mean, it may be that way, but I think that somehow the electron is represented by the wave function, which is just a probability, and if we want to localize it then we lose some of the information.  So whether this is true or not is something of a philosophical question. I wish I knew, or understood it, but I don't. For now, for me, the electron is the wave function, so whether the electron is distributed among the wave function, and when you do an experiment, it sucks into one point, or whether it is indeed one particle at a point, statistically the average, I don't know.\end{quote}

We characterize Student J's agnosticism as sophisticated because he recognizes there are multiple ways of thinking about the situation, but doesn't yet have sufficient information to decide for himself which is more likely to be correct.  This sophistication might not be apparent in other students who profess agnosticism, which can in some cases actually be more of an expression of confusion as to what it is that's going on in the first place.

\section{\label{sec:Conclusion}Discussion and Instructional Implications}

The preceding sections include multiple examples of students and experts employing epistemological resources that influence their ontological categorization of electrons as particles: localized detections imply a continuously localized existence; massive (or charged) particles must be localized in space; particles are by definition localized.  These types of resources are not necessarily wrong, and might be productive when working with classical systems.  However, they can have important implications for what kind of physical meaning is attached to the otherwise algorithmic process of deriving wave functions and calculating expectation values.  They may also interfere with students attaining a deeper understanding of quantum mechanics.

While some of the interviewed students did show a lack of flexibility in their conceptions of electrons, most others exhibited ontological flexibility to varying degrees: some distinguished between intuitive and normative ontologies; some perceived switches between particle and wave descriptions as physical transitions; others blended attributes from classically distinct categories, or assigned them separately according to which would be most productive in that specific context.  We believe this type of bridging between parallel ontologies should be promoted during instruction because it is an important aspect of learning and understanding quantum mechanics.

Moreover, if a single normative (scientifically accepted) ontology for quantum entities does not exist, then Chi and Slotta's strategy of `ontological correctness' during instruction simply cannot be applied.  There is no consensus among practicing physics regarding the ``correct'' ontological nature of electrons; in fact, most actually do not concern themselves with such issues when engaged in their research.  This reality is reflected in the variety of approaches instructors take when addressing interpretive themes, often de-emphasizing the physical interpretation of quantum theory in favor of developing mathematical tools, with demonstrable impact on student thinking \cite{baily2015modern}.  We would argue that instructors who take a Copenhagen/Agnostic approach to teaching quantum mechanics are in essence preventing students from developing mental models of quantum processes that could be productive for their understanding.  We have also demonstrated that students will develop these mental pictures on their own in contexts where instruction is less explicit, and that such ideas have a greater tendency to be intuitively classical, so that directly confronting these intuitions in multiple contexts is essential.

Nor should we connote too much negativity with students relying on their intuition as a form of sense making.  It is true that ``everyday thinking'' can sometimes be misleading in quantum physics, but that is not a sufficient argument for the complete abandonment of tools that still have a range of applicability.  It is not that a strictly particle view of electrons is illegitimate, but the consistent application of this ontology in all contexts can lead novices to paradoxical or incorrect conclusions.  Ballentine may be content with describing diffraction patterns in terms of quantized momentum transfers, but this argument entails concepts and mathematical tools that lie outside the scope of an introductory course.  It is also unclear how this view can be reconciled with the fact that atomic electrons do not radiate in the ground state, or that the orbital angular momentum of a ground state electron is zero in the Schr\"{o}dinger model of hydrogen.   We would argue that one goal of instruction should be to help make clear for students precisely when thinking about an electron as a particle might lead them astray, and when it would not (and similarly for waves).

\begin{figure}
\begin{framed}
    \includegraphics[height=1.5in, width=3in]{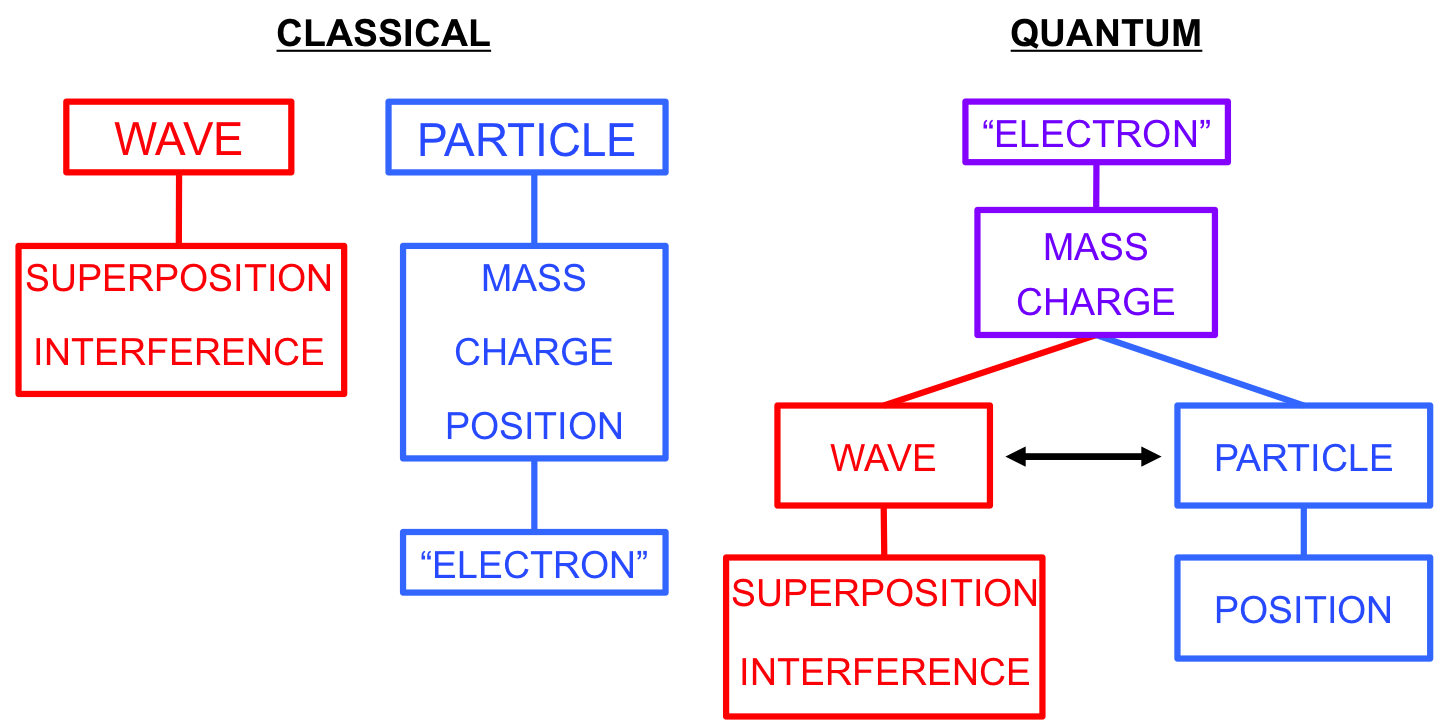}
\end{framed}
\caption{A comparison of classical and quantum ontologies, wherein attributes such as mass and charge that are typically associated with classical particles are separated out to become a more global property of quantum electrons, while wave and particle attributes are relegated to separate sub-categories.}\label{fig:ontology}
\end{figure}

At the very least, we would recommend that instructors be explicit with students about consciously separating classical and quantum ontologies (as in Fig. \ref{fig:ontology}), so that some of the reasoning we see invoked by students (i.e., classical attribute inheritance) might be avoided.  However, we believe it makes more sense for instructors to appeal to students' intuitions about classical particles and waves when teaching quantum mechanics (i.e., leverage students' prior knowledge), while at the same time developing epistemological tools to aid students in deciding when either of these descriptions is most appropriate.

Informed by the results of our research, we have developed a modern physics curriculum with multiple aims, among them: (i) make the physical interpretation of quantum physics a course topic unto itself, and consistently attend to student ontologies throughout the course; (ii) help students acquire the language and resources to identify and articulate their own (often unconscious) beliefs about reality and the nature of science; and (iii) provide experimental evidence that directly confronts their intuitive expectations.  Our ultimate goal was for students to be able to distinguish between competing points of view, to recognize the advantages and limitations of each, and to apply this knowledge in novel situations.  In short, instead of trying to tell students what they should and shouldn't believe about quantum physics, we provided them with logical arguments and experimental evidence, then let them decide for themselves.  This modern physics curriculum has been implemented twice now at the University of Colorado Boulder, with results that indicate many of our objectives were achieved, some of which are reported in Ref. \cite{baily2015modern}, and in greater detail in Ref. \cite{baily2011thesis}.

\begin{figure}
\begin{framed}
    \includegraphics[height=1.5in, width=3in]{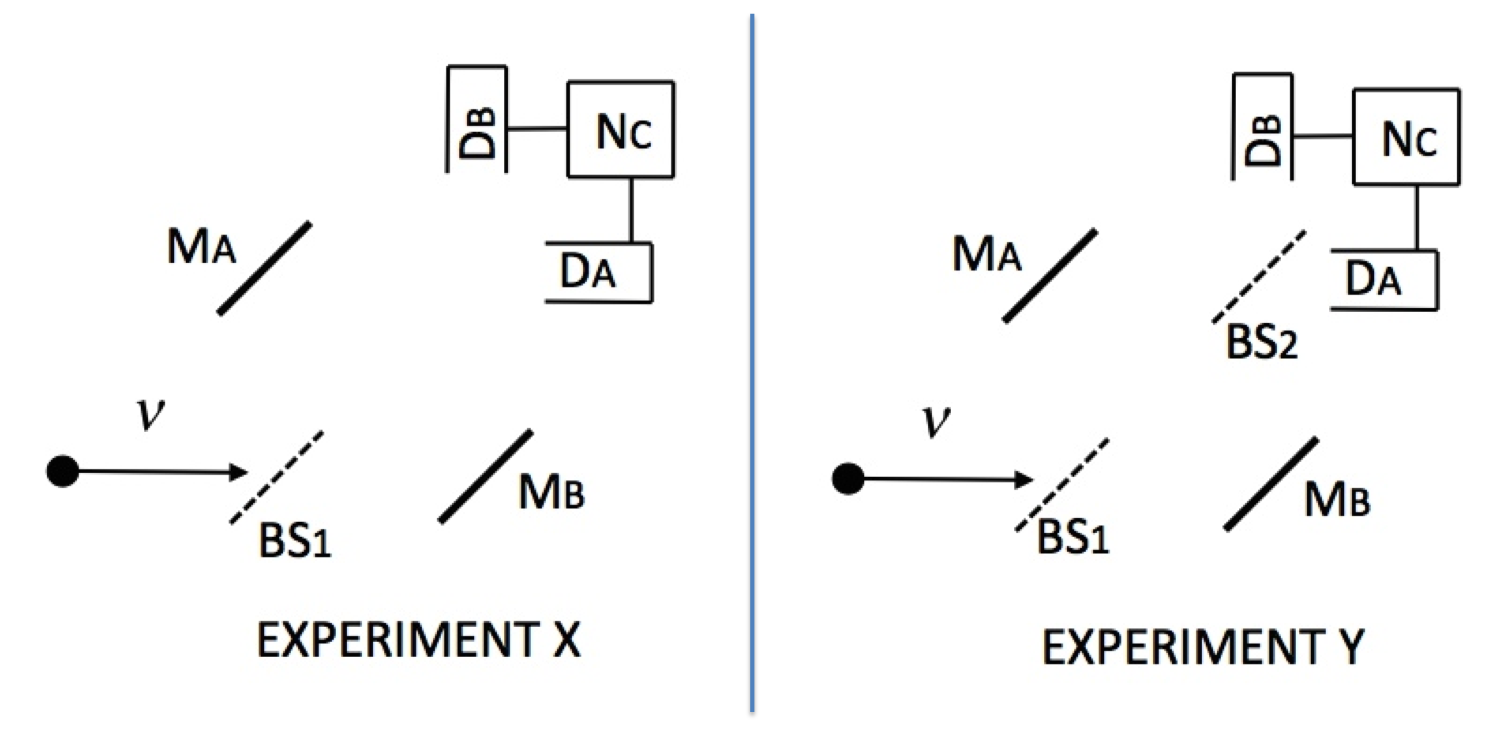}
\end{framed}
\caption{In each of these two experiments X (one beam splitter) \& Y (two beamsplitters), a single photon ($\nu$) is sent to the right through the apparatus. M = Mirror, BS = Beam Splitter, D = Detector, N$_{C}$ = Coincidence Counter.}\label{fig:experimentxy}
\end{figure}

We decided to promote a Matter-Wave interpretation in this course, because we believe it provides students with the most consistent way of interpreting quantum phenomena.  For example, one topic in this curriculum was single-photon experiments with a Mach-Zehnder interferometer \cite{grangier1986photon}. [See Fig. \ref{fig:experimentxy}.]  When just a single beam splitter is present (Experiment X), each photon is recorded in either one detector or the other, but never both simultaneously {-} this result is often interpreted as meaning each photon takes just one of the two paths with 50/50 probability.  When a second beam splitter is present (Experiment Y), interference effects can be observed by modulating the path length in just one of the arms of the interferometer, which many interpret as each photon taking both paths simultaneously.  In a delayed-choice experiment \cite{jacques2007delay}, the second beam splitter is either inserted or removed \textit{after} the photon has encountered the first beam splitter, but while it is still traveling through the apparatus.  Interference is observed if the second beam splitter had been inserted, and otherwise not.

Some instructors would teach that whether the second beam splitter was present or not determines whether the photon took both paths or just one.  This explanation is dubious in light of how the choice between configurations takes place outside the light cone of the photon's encounter with the first beam splitter; and students may question how the photons (or they themselves) are to know which type of behavior should occur.  We taught our students that each photon always takes both paths simultaneously, regardless of whether the second beam splitter is present or not.  On the other hand, we also explained that students could decide which type of behavior to expect depending on the ``path information'' available.  If it could be determined which path a photon had taken, then interference would not be visible; if not, then interference effects could be observed.  In doing so, we appealed to students' intuitions about classical particles (they are either reflected or transmitted) and classical waves (they are both reflected and transmitted).  [Note that these strategies may also be employed in the context of the double-slit experiment.]

In a midterm exam question during the first implementation of this curriculum, students were asked to explain in which experiment (X or Y) they would expect photons to exhibit wave-like behavior.  Nearly 40\% of students focused on the lack of which-path information, and 33\% said the availability of two paths for the photon was key to predicting wave-like behavior in Experiment Y.  As an illustrative example, one student explained in their exam response: ``Since [the photon] can take either path and still get to either photomultiplier, I know it can be represented as a wave [in Experiment Y].''

There is still a great deal of research to be done regarding students' quantum ontologies, but we have attempted with this paper to lay some of the groundwork.  Considering that we have observed strong associations for students between particles and definite paths, it seems that curricular development efforts should explore more deeply the usefulness for students of `which-path information' as an epistemological tool [two paths = interference (wave); one path = no interference (particle)].

It would also be worthwhile to explore what connections (if any) exist between the ontologies students employ and their ability to set up and solve quantum mechanics problems; also, whether there are specific contexts in which one ontology has definite advantages over others in terms of quantum mechanical calculations.  In particular, we suspect that many of the known student difficulties with quantum measurement \cite{zhu2012qm1,zhu2012qm2} are rooted in their lack of mental representations of the actual measurement process and corresponding reduction of the quantum state.

We also recognize that very little is understood about the thinking associated with Agnostic students.  Sophisticated agnosticism would acknowledge the existence of evidence that favors more than just a single interpretation; at the same time, an Agnostic stance may be indicative of the perception that nothing can truly be known or understood in science.  Therefore, a deeper exploration is called for of how students' views on the nature of science are impacted by learning about quantum physics.

\begin{acknowledgments}

This work was supported in part by NSF CAREER Grant \# 0448176, NSF DUE \# 1322734, the University of Colorado and the University of St Andrews.  Particular thanks go to the students and instructors who made these studies possible.

\end{acknowledgments}

\bibliography{references}
\bibliographystyle{apsper}   

\end{document}